\documentclass[a4paper,11pt]{article} 

\newif\ifINCLUDENOT   

\INCLUDENOTtrue       
 \INCLUDENOTfalse    

\newif\ifINCLUDE   
\INCLUDEtrue       

\ifINCLUDENOT

Possible Journals

PLOS

ELIFE
https://sheffield.ac.uk/library/library-resources-collections/content-strategy/library-support-equitable-oa#c
Our support contributes to eLife, a not-for-profit organization that publishes open access research in the life and biomedical sciences, aiming to improve the experience of publishing and evaluating research.

Known for being open to "Re-evaluations" or "Formal Analysis" of existing high-profile datasets. They favor papers that offer a clean, elegant explanation for complex biological phenomena.

\fi

\RequirePackage{fullpage}
\RequirePackage{amsmath}
\RequirePackage{graphicx}
\RequirePackage{setspace}
\RequirePackage{latexsym}
\RequirePackage{amsfonts}
\RequirePackage{epsf}
\RequirePackage{url}
\RequirePackage{subfig} 
\RequirePackage{ulem}
\RequirePackage{soul}
\RequirePackage{amsthm}
\RequirePackage{thmtools}
\RequirePackage{setspace,lipsum}

\renewcommand{\em}{\it}

\renewcommand{\ni}{\noindent}

\newcommand{\be}        { \begin{equation}  }
\newcommand{\ee}        { \end{equation}	}
\newcommand{\bea}       { \begin{eqnarray}  }
\newcommand{\eea}       { \end{eqnarray}    }

\newcommand{\E}{\mathrm{E}}

\newcommand{\bits}	{\text{ bits}}

\newcommand{\kappA}{\beta}

\begin{document}
\onehalfspace


\thispagestyle{empty}

\title{{\bf Efficient Coding Predicts Synaptic Conductance  }}


\author{
James V Stone, 
University of Sheffield, England. 
j.v.stone@sheffield.ac.uk\\
File: arxiv\_synapseJVStone2026\_v9withcorrection.tex, \\ 
ARXIV: \url{https://arxiv.org/abs/2603.03347}
}

\date{\today}
\maketitle

\newcommand{\s}{\sigma}

\begin{abstract}
Synapses are information efficient in the sense that their natural conductance values convey as many bits per Joule as possible, but efficiency falls rapidly if the conductance is forced to deviate from its natural value (Harris et al, 2015). However, the exact manner in which efficiency falls as conductance deviates from its natural value remains unexplained. Recently, Malkin et al (2026)  showed that synaptic noise is minimised given the available energy, consistent with a minimal energy boundary.  This minimal energy boundary is a necessary, but not sufficient, condition for maximising information efficiency. By expressing the minimal energy boundary in terms of Shannon's information theory (Shannon, 1949), we show that synapses operate at signal-to-noise ratios which maximise information efficiency, and that this accurately predicts the decrease in efficiency values observed in Harris et al (2015)  across a wide range of synaptic conductances. Crucially, the proposed model contains no free parameters because it is derived from the biophysics of the synapse. The results reported here are consistent with the general principle that neuronal systems in the brain have evolved to be as efficient as possible in terms of the number of bits per Joule.
\end{abstract}

\section*{Introduction}

\newcommand{\eff}{\varepsilon}

Harris et al  (2015)\nocite{harris2015} provide support for efficient coding in terms of bits per Joule. In their study, the conductances of synapses between the optic nerve and the {lateral geniculate nucleus}  were artificially altered from their natural values, and the (mutual) information transmitted per Joule expended on synaptic transfer was measured at each conductance value.   
It was found that each synapse transmits most information per Joule at its natural conductance, as shown by the data points in Figure \ref{harris2015a}.

\begin{figure}[b!] 
\begin{center}
\includegraphics[ width =0.7 \textwidth, angle=0 ] {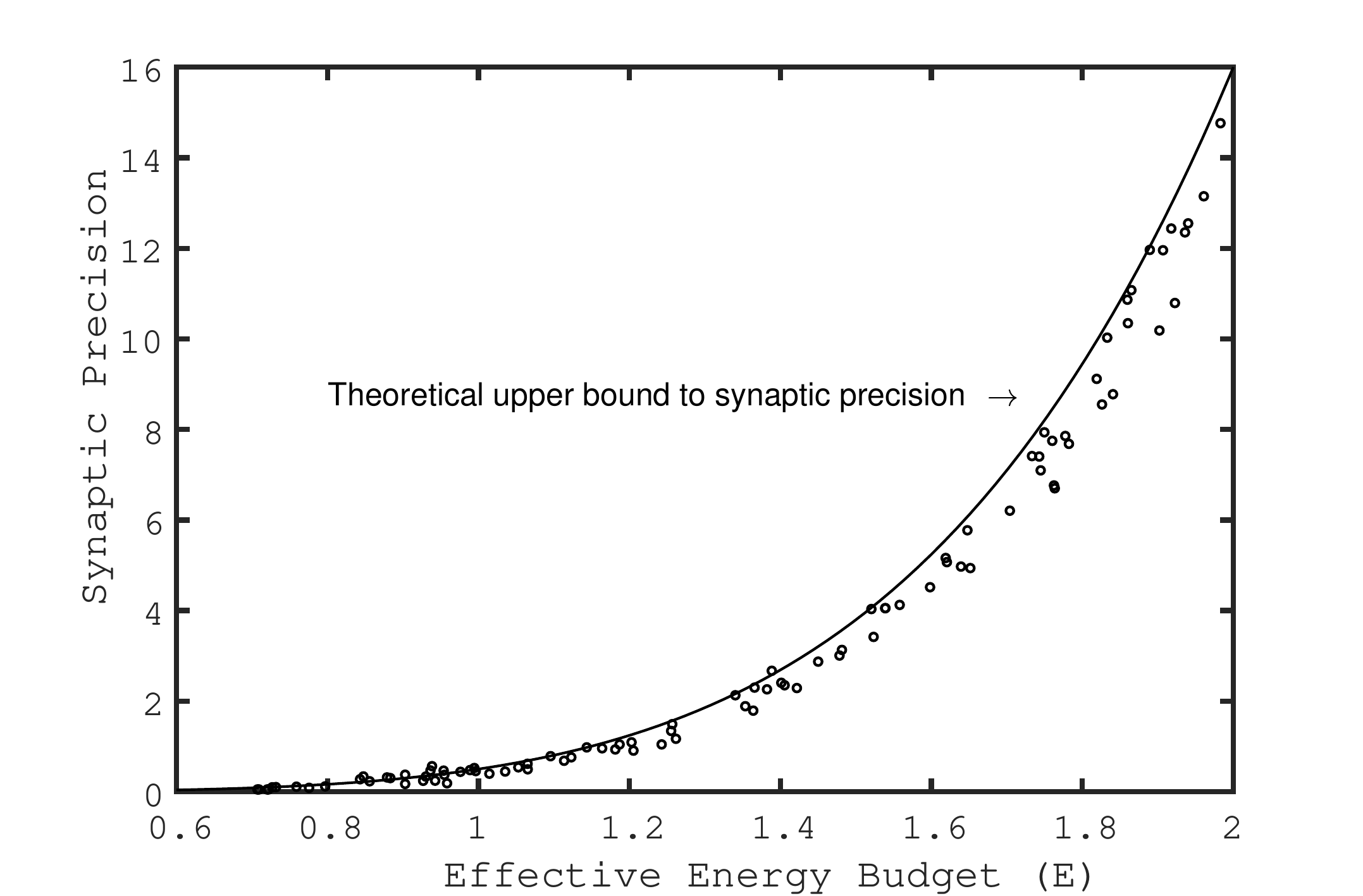}
\caption{How precision $\sigma^{-2}$ increases with energy budget $E$ at a synapse.  
The curve is the theoretical upper bound to precision (Equation \ref{SNRu}, $k$=$0.5$, $\mu$=1). 
Units are arbitrary and data points are schematic. 
}
\label{synapseenergy20261} 
\end{center}
\end{figure}

%
Recently, Malkin et al (2026)\nocite{malkin2026energybudget} demonstrated that neuronal data sets from five published studies conform to a minimal energy boundary condition, such that the energy budget determines an upper bound to synaptic noise precision  (Figure \ref{synapseenergy20261}). 
Their paper is summarised next.

For any given synaptic conductance, it is important that the variance of post-synaptic potentials is as small as possible. 
However, this variance ultimately depends upon the diffusion of neurotransmitter vesicles across the synapse, which is inherently noisy.  
Thus, even for a synapse with an optimal synaptic conductance, energy is wasted if key physiological parameters have values that 
do not minimise  the noise variance in post-synaptic potentials, which corresponds to maximum precision. 

These physiological parameters are: 
1) the number $n$ of number of pre-synaptic  active zones where neurotransmitter vesicles can be released, 
 2) the probability $p$ that a vesicle will  be released from any given site when a presynaptic  spike arrives, 
and, 3)  the size $q$ of the electrical response produced in the post-synaptic neuron by a single vesicle. Using binomial assumptions, the mean evoked post-synaptic  potential is 
\bea
	\mu & = &  npq, \label{eqmuGmalkin}
\eea
 and the noise variance (across different trials) in the evoked post-synaptic potential is
\bea
	\sigma^{2} & = & np(1-p)q^{2}\\
	 & = & \mu (1-p)q.
\eea
Malkin et al (2026) defines precision as the signal-to-noise ratio, which allows synapses of different mean efficacies to be compared,
\bea
	SNR & = & \mu^{2} / \sigma^{2}. \label{SNRyt}
\eea
A key result of Malkin's analysis defines  a minimal energy boundary (in their Equation 20), which predicts that the signal-to-noise ratio increases in proportion to the 5th power of the  energy budget $E$ available at a synapse. Specifically, 
\bea
	 \mu^{2} / \sigma^{2} & = & k E^{5},  \label{SNRu}
\eea
where $k$ is constant for a given synapse, and its value determines the balance between energy expended on presynaptic calcium pumps and energy expended on neurotransmitter vesicle maintenance. 
By analysing  data from five published studies on synaptic plasticity, Malkin et al (2026) found that synapses operate close to the minimal energy boundary. 

\section*{Predicting Synaptic Efficiency}
\begin{figure}[t!] 
\vspace{-0.2in}
\begin{center} 
\subfloat[]{\includegraphics[width=0.5\textwidth] {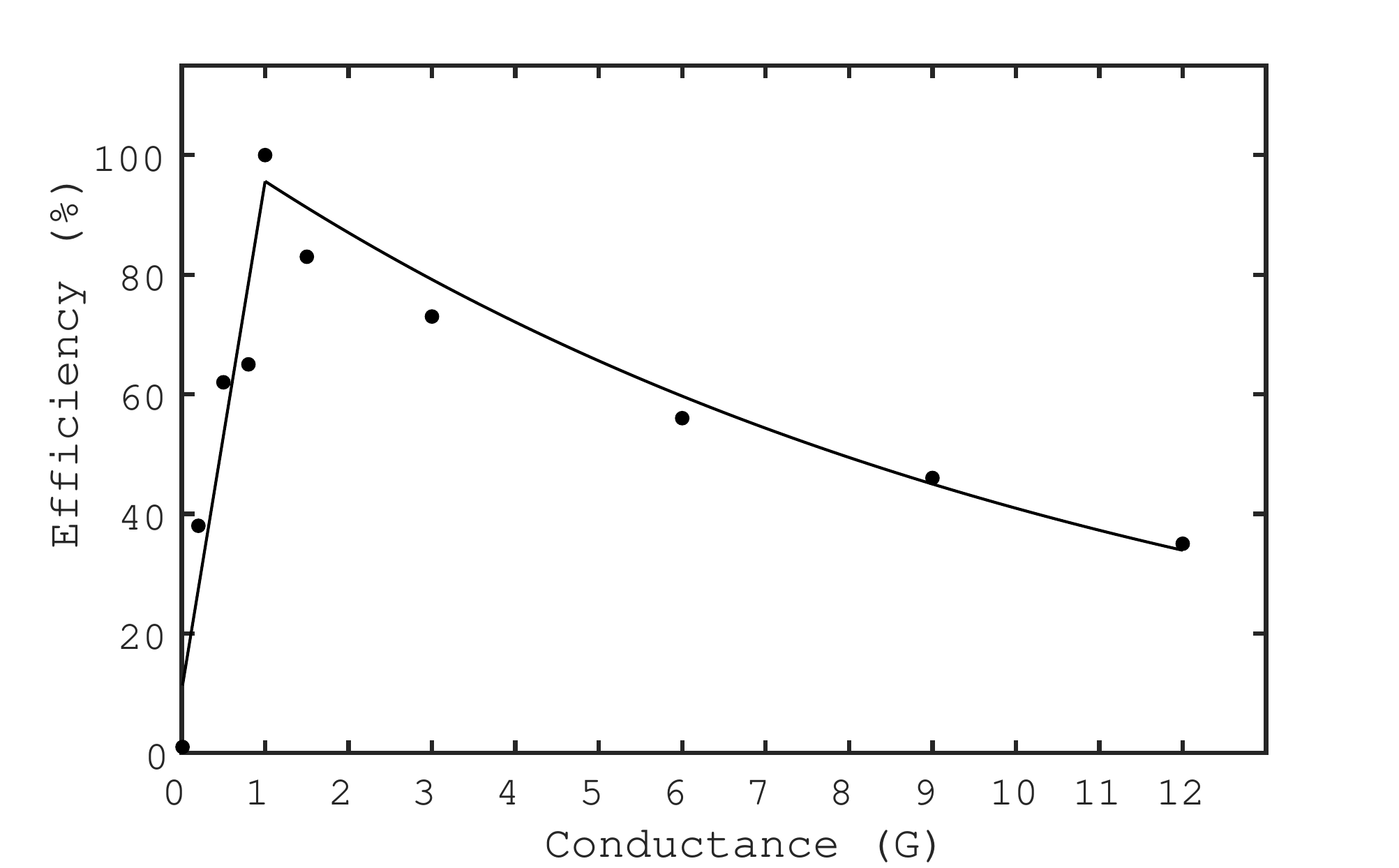} }
\subfloat[]
{\includegraphics[width=0.5\textwidth] {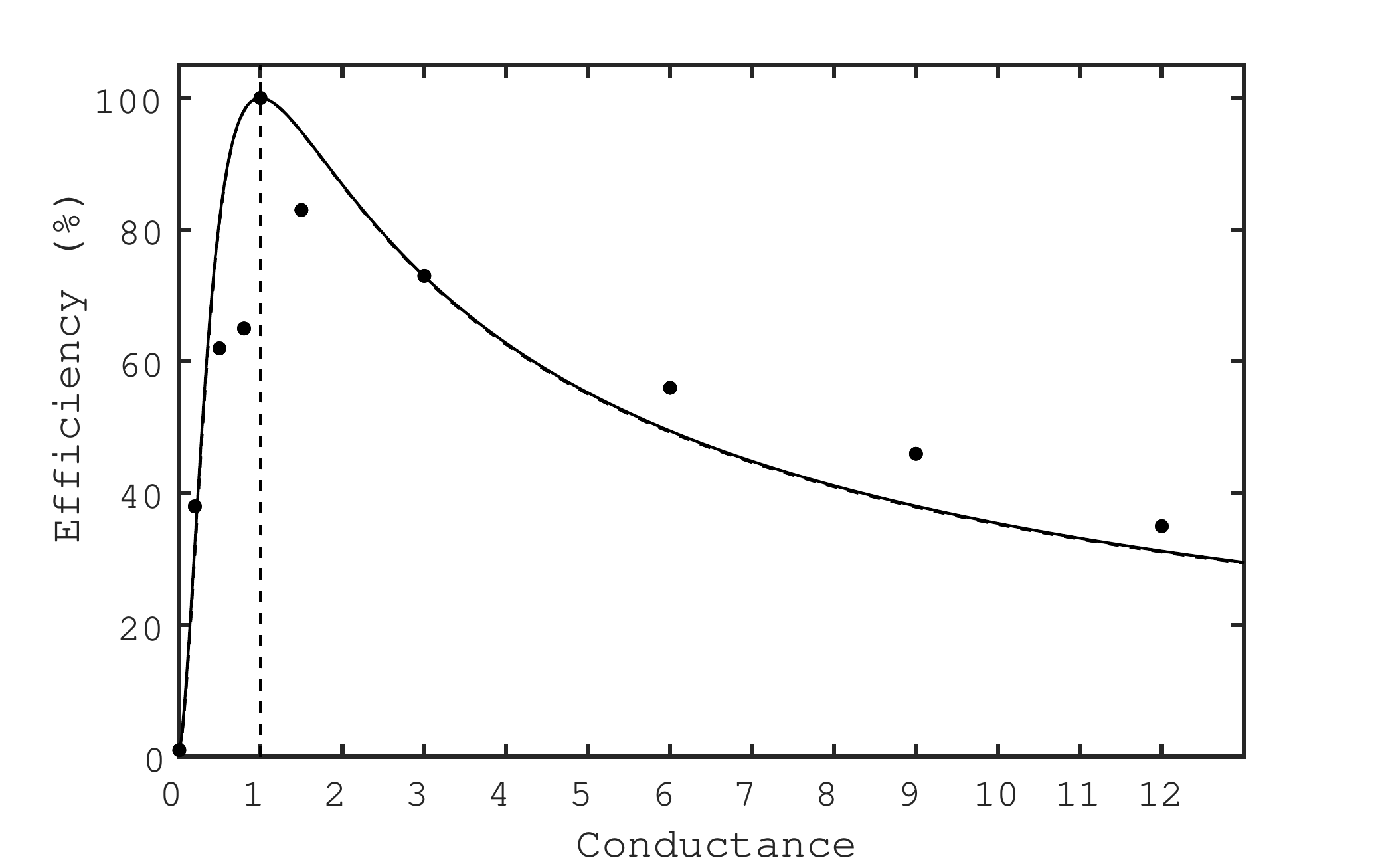} }
\caption{
 Maximum efficiency occurs at the natural mean synaptic conductance value.\\
(a) Harris et al  (2015) fitted a line for mean conductance values $<1$, and fitted an exponential for mean conductance  values $>1$. Data points are experimentally measured values of mean conductance, estimated from Figure 5E in Harris et al  (2015).  Efficiency values have been re-scaled to percentages of efficiency at the natural mean conductance. 
\\ (b)  
 The dashed curve uses the theoretically optimal value of  $\alpha$=2.5 in Equation \ref{eqsd5deabA}. 
 The (almost identical)  solid curve is Equation \ref{eqsd5deabA} fitted  to data from Harris et al  (2015), which yields $\hat{\alpha}$=2.451. 
 Data points are experimentally measured values of mean conductance, estimated from Figure 5E in Harris et al  (2015).  Efficiency values have been re-scaled to percentages of efficiency at the natural mean conductance. 
}
\label{harris2015a}
\end{center}
\vspace{-0.2in}
\end{figure}

%
A scatterplot of conductance $G$ versus efficiency $\eff$ from Harris et al  (2015) is shown in Figure \ref{harris2015a}. 
In Harris et al  (2015), purely for convenience, efficiency values were fitted to a line for conductance values $G<1$, and fitted to an exponential function for conductance  values $G>1$.  
Here, we express the minimal energy boundary proposed in Malkin et al (2026) in terms of Shannon's information theory (Shannon, 1949) to derive a 
model that accurately  predicts  the decrease in efficiency values plotted in Figure \ref{harris2015a}. 

For a synapse which uses $E$ Joules to convey  $I$ bits, we define information efficiency as
\bea
	\eff & = &  {I} / { E } \: \text{ bits/Joule}. \label{eqsddf45ab}
\eea
To derive the desired parametric function,  mutual information $I$ and energy $E$ need to be re-expressed in terms of synaptic conductance.   
First,  Shannon's equation\cite{ShannonWeaverBook} for mutual information needs to be re-expressed. 
Shannon defined mutual information as
\bea
	I 	& = &  \log \left( 1 + SNR \right)^{ {1}/{2}}  \label{eqshannonA} \\
		& = &  \log \left( 1 + \frac{\mu^{2}} {\sigma^{2}} \right)^{ {1}/{2}} \bits. \label{eqshannon}
\eea
 Substituting Equation \ref{SNRu} into Equation \ref{eqshannon} yields 
\bea
	I & = &   \log \left( 1 +    k E^{5} \right)^{ {1}/{2} } \bits.
\eea
This implies that, if synapses operate close to the minimal energy boundary then
 a small increase in available energy $E$ provides a relatively large increase in SNR, and a correspondingly large increase in information. 

Ohm's law dictates that the energy expended on the noise variance in the  evoked post-synaptic  potential $V_{noise}$ is 
\bea
	E & = & V_{noise}^{2}G_{1}  \text{ Joules}. \label{eqetransa}
\eea
In Harris et al  (2015), $G_{1}=\E[\mu]$ is the mean conductance of the synapse, where this expectation was taken over 10 cells, each of which was recorded for 125 seconds, and normalised conductance was defined as
\bea
	G & = &  \frac {\text{artificially induced conductance}} {\text{natural conductance}} = \frac{G_{1}}{G_{0}},
\eea
 so $G$ is measured in units of the synapse's natural conductance, where the natural conductance was observed to maximise $\eff$, which corresponds to $G=1$.  
 Given that $G_{1} =   G_{0} G$, the energy expended on the average (RMS) noise voltage $V_{noise}$ is
\bea
	E 	& = & V_{noise}^{2}G_{0} G \\
		 & = & \kappA G  \text{ Joules},   \label{eqetrans} 
\eea
where the energy constant is $\kappA =  V_{noise}^{2}G_{0} \text{ Joules}$. 
Therefore, the mutual information is
\bea
	I & = &   \log \left( 1 +   k (\kappA G)^{5} \right)^{ {1}/{2}} \bits, \label{ffigere}
\eea
where the signal-to-noise ratio SNR=$ k (\kappA G)^{5}$. 
If the SNR is large then 
\bea
	\log (1+SNR)^{1/2} & \approx & \log (1+SNR^{1/2}),
\eea
 so
\bea
	I
	 & \approx &    \log \left(  1 +   k^{1/2}  \kappA^{2.5} G^{2.5} \right) \\
	 & = &   \log \left(1+  c G^{\alpha} \right)  \bits,  \label{ffigeread}   \label{eqCby56}
\eea
where the exponent $2.5$ has been replaced with the parameter $\alpha$ in the final equation, and 
where the constant
\bea
	c & = &   k^{1/2}  \kappA^{\alpha}.    \label{ffigereadc}
\eea
According to Harris et al (2015), the data in Figure \ref{harris2015a} has $4.7$ bits per spike, which refers to the amount of mutual information  transmitted through a synapse by each spike. A mutual information of $4.7$ bits implies a signal-to-noise ratio of SNR $\approx  675$ (Equation \ref{eqshannon}); so the assumption of a large signal-to-noise is justified here. 

Second, we express energy in terms of conductance, which has been done in Equation \ref{eqetrans}.
 Substituting Equations \ref{eqetrans}  and \ref{eqCby56}   in Equation \ref{eqsddf45ab}, 
defines the information efficiency 
\bea
	\eff & = &   \frac{\log (1 + c G^{\alpha})}{ \kappA G } \: \text{ bits/Joule},  \label{eqsd5deab}   \label{ffigereadcA}  \label{eqsd5deabA}
\eea
where the value of $c$ is uniquely determined given that $\partial \varepsilon/\partial G=0$ at $G=1$. 

We find the value of $\alpha$ that maximises $\eff$ as follows. 
For each putative value of $\alpha$, the optimisation process involves a numerical search for the value of $c$ that minimises the sum of squared differences between $\eff(\alpha)$ (Equation \ref{eqsd5deabA}) and observed values of $\eff$.
At each value of $\alpha$, we ensure that the peak value of $\eff$ is at $G=1$. Equivalently, we set the value of $c$ to ensure that ${\partial \varepsilon}/{\partial G}=0$ at $G=1$, where
\bea 
	\frac{\partial \varepsilon}{\partial G} & = & \frac
									{\left( \frac{\alpha c G^{\alpha}}  {1 + c G^{\alpha}} \right)  
									- \log(1 + c G^{\alpha})} 
										{ \kappA G^{2}}
\eea
If we set this to zero and set $G=1$ then we have
\bea
	\frac{\alpha c}{1 + c} & = & \log(1 + c). \label{eqac}
\eea
For a given value of $\alpha$, we solve for $c$ numerically to satisfy the requirement that $\eff$ is maximised when $G=1$.

\vspace{-0.2in}
\section*{Results}
\vspace{-0.1in}
The  model was evaluated using experimental data from Harris et al  (2015), shown as data points in Figure \ref{harris2015a}. 
Using the information efficiency function (Equation \ref{eqsd5deabA}) with $\kappA=1$ and the  theoretically derived value $\alpha=2.5$ (for which the value $c= 8.315$ ensures the peak value of efficiency $\eff$ is at $G=1$), a goodness-of-fit $F$-test yields $R^2= 0.7496$ with a $p$-value of  $p=0.0012$ ($F(1,8)= 23.95$). This function (Equation \ref{eqsd5deab}) is shown by the dashed curve  in Figure \ref{harris2015a}b.

Next, we ran a goodness-of-fit test to see if a value of $\alpha$ different from the  theoretically derived value $\alpha=2.5$ would provide a better fit to the data. A numerical search for the value of $\alpha$ that provides the best least-squares fit of Equation \ref{eqsd5deabA}, 
yielded $\hat{\alpha}=2.453$, for which the value 
$c(\alpha)=7.796$ ensures the peak value of efficiency $\eff$ is at $G=1$. A goodness-of-fit  $F$-test yielded 
$p=0.0078$ ($F(1,7)=10.490$, $R^{2}=0.7498$). 
The fitted function is shown by the solid curve in Figure \ref{harris2015a}b (which is almost identical to the dashed curve for which $\alpha=2.5$).

The value $\alpha=2.5$ is based on the assumption that  synapses operate at the minimal energy boundary. 
If this assumption were not valid then obtaining a value for $c$ which ensures  the efficiency maximum occurs at $G=1$ would result in a curve in Figure \ref{harris2015a}b that is inconsistent with the data. 
Thus, aside from the biophysically motivated estimate of $c$ (used to ensure efficiency peaks at $G=1$, as observed experimentally), the model represented by Equation \ref{eqsd5deab} contains no free parameters. 
\vspace{-0.2in}
%
\section*{Discussion} 
\vspace{-0.2in}
%
Shannon's information theory dictates that the energy cost of information rises steeply with the rate at which information is transmitted, which suggests that neurons within the brain should process information at the lowest energy cost.   
By expressing information efficiency in terms of synaptic conductance, it has been shown here that the particular rate at which efficiency falls as synaptic conductance deviates from its natural value is consistent with the minimal energy boundary established by Malkin et al (2026). 

Even though the minimal energy boundary demonstrates that synaptic noise variance $\s^{2}$ is minimised for a given energy budget, it does not guarantee that information transfer $I$ is maximised for a given energy budget; and therefore the minimal energy boundary  does not guarantee that efficiency $\eff$ is maximised. 
Thus, from an information-theoretic perspective, the  the minimal energy boundary defines a necessary, but not sufficient, condition for maximising efficiency. 
For example, if some of the total energy associated with a synapse is wasted then the signal variance would be reduced,
which would reduce the  mutual information $I$, and therefore  efficiency $\eff$ would be non-maximal, regardless of the noise variance $\s^{2}$. 
Thus, whereas Malkin et al (2026)  showed that synapses minimise noise variance, results presented here demonstrate that synapses also operate at specific signal-to-noise ratios which ensure information efficiency is maximised across a wide range of conductances. 

Crucially, the new efficiency function  (Equation \ref{eqsd5deab}) has no free parameters, so the efficiency function plotted in Figure \ref{harris2015a} has not been fitted to the data from Harris et al (2015).
This is consistent with the general principle
 that neuronal systems in the brain have evolved to be as efficient as possible in terms of the number of bits per Joule.


\ni
{\bf Acknowledgment}: Thanks to Raymond Lister for comments on a draft of this paper.



\end{document}